\documentclass[aps,prd,reprint,superscriptaddress,showpacs,showkeys,amsmath,amssymb,floatfix,nofootinbib]{revtex4-1}

\usepackage{verbatim}
\usepackage{graphicx}
\usepackage{dcolumn}
\usepackage{bm}

\begin{document}
\preprint{APS/123-QED}

\title{A test of reality using quantum entanglement in $J/\psi \rightarrow \Lambda\bar{\Lambda}$}

\author{Xinyan Tong}
\email{txyz$\_$cc@pku.edu.cn}
\author{Qiaorong Shen}%
\email{shenqr@pku.edu.cn}
\author{Yunfei Long}
\author{Yajun Mao}
\affiliation{%
School of Physics and State Key Laboratory of Nuclear Physics and Technology, Peking University
}%

\date{\today}

\begin{abstract}
Quantum mechanics (QM) has been criticized for its violation of locality and reality since it was put forward. A series of hidden variable theories (HVTs) were evolved to ensure reality. In this paper, a new method to distinguish QM and HVTs is proposed. The angle between two decay planes is introduced as the observable and the distribution of it under two conditions is deduced. The feasibility to test it on BESIII experiment is discussed.
\begin{description}
\item[PACS numbers]
03.65.Ud,13.25.Gv
\end{description}
\end{abstract}

\maketitle



Quantum mechanics (QM) has been precisely tested by a number of experiments. However, there are still heated debates about the interpretation of QM. In 1935, Einstein, Podolsky, and Rosen (EPR) argued that QM is incomplete for its violation of locality and reality \cite{EPR}, which is well known as the EPR paradox. In 1951, Bohm rephrased the paradox in a clearer way \cite{Bohm_reexpression} with two spin-1/2 fermions in a system of total-spin singlet. A series of substitute theories called local hidden variable theory (LHVTs) were proposed to escape the paradox \cite{Bohm_theory}.

In 1965, Bell constructed mathematical inequalities \cite{Bell_inequality} through which experimental tests could distinguish between QM and LHVTs by testing locality. Many experiments, mostly using the polarization correlation of entangled photon pairs, have been performed since then whose results showed consistency with QM instead of LHVTs \cite{entangled_photon}.

Later, some non-local hidden variable theories were evolved mainly by Bohm and De Broglie, which proposed that the realism of nature refers to ``objective reality'' instead of ``contextual reality'' \cite{landr}. The former means that the observed property is pre-exist before measurement, while the latter indicates that observed property arise only after observation. The latter interpretation is suggested by QM and violates reality. Afterwards, Leggett brought up another inequality \cite{Leggett} to test the reality of the non-local hidden variable theories, which was also tested by entangled photon pair experiments \cite{2007}.

Apart from utilizing the laser beam, the validation of QM has also been carried out by the high energy physics experiment, e.g.\ $B_0\overline{B_0}$ system \cite{bb}, utilizing the coincidence rate of two B mesons' flavor asymmetry to construct the Bell's inequality, where the violation of the inequality has been seen. 

In the experiment of high energy physics, the angular distributions are feasibly obtained. With the angular distributions of decay, T$\mathrm{\ddot{o}}$rnqvist suggested that weakly decay reaction like $e^+e^+\to\Lambda\bar{\Lambda}\to p\pi^-\bar{p}\pi^+$ could be used to test the non-locality of QM \cite{torn}, which has been performed by DM2 Collaboration \cite{DM2}. However, due to the insufficient statistics, they couldn't draw a decisive conclusion. BESIII is also regarded as an ideal high energy facility to test the Bell's inequality, utilizing the entanglement of vector meson pairs \cite{chen}\cite{lxq}\cite{xzj}. 

Here we use the angular distribution directly to test the nature of reality instead of applying the inequalities. In this paper, the decay chain \[J/\psi\to\Lambda\bar{\Lambda}\to p\pi^-\bar{p}\pi^+\] is analyzed as a detailed description. The distribution of the angle between the two decay planes of $\Lambda$ pair, denoted as $\alpha$, is taken as the observable. We expect the existence of entanglement between $\Lambda$ and $\bar{\Lambda}$ will reflect difference on the distribution. Whether $\Lambda$ and $\bar{\Lambda}$ are entangled or not refers to ``contextual reality'' or ``objective reality'' respectively. Hence, the distribution of $\alpha$ is supposed to distinguish the two interpretations of reality.



\begin{figure}[h]
\centering 
\includegraphics[width=9.5cm]{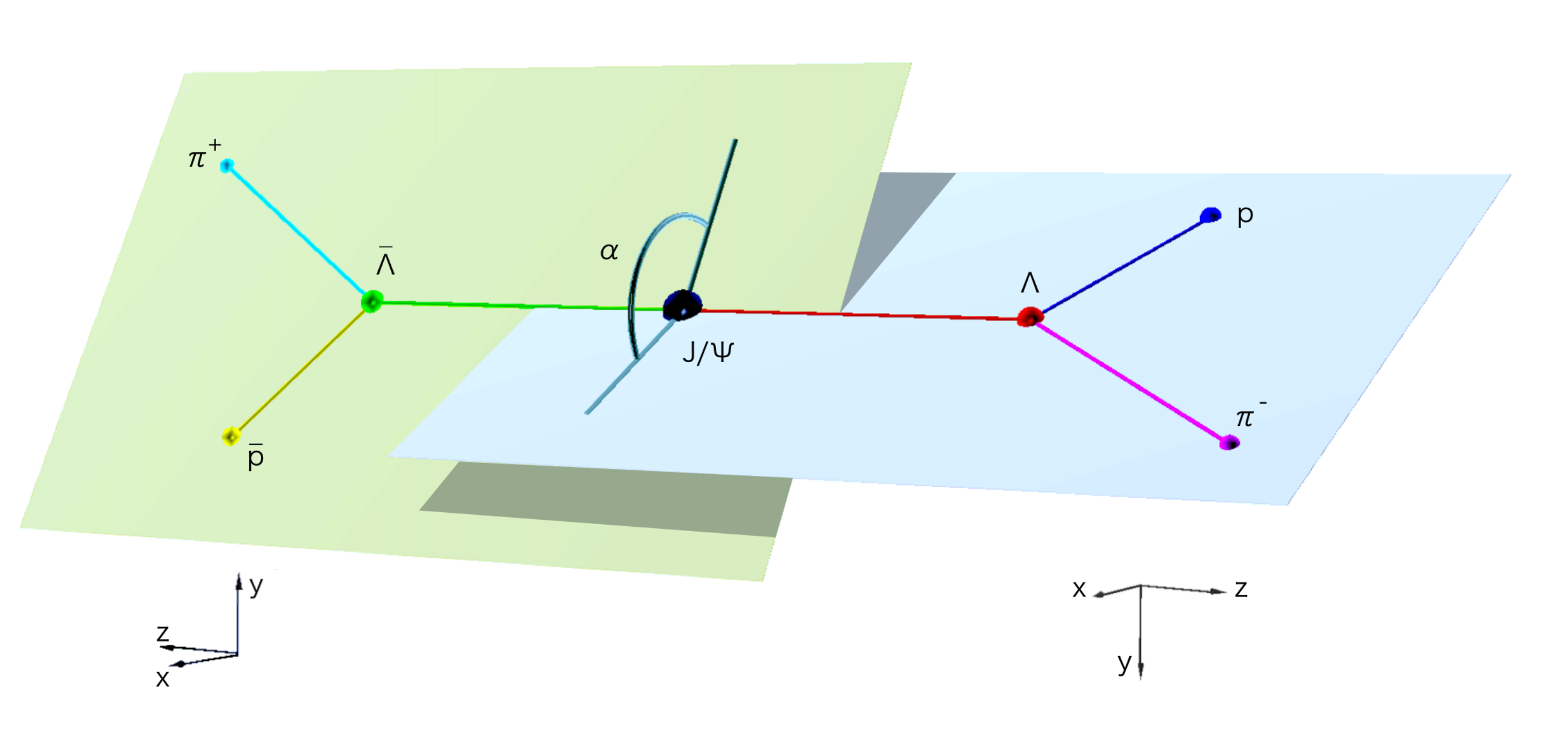}
\caption{\label{fig:plane}%
 $J/\psi\to\Lambda\bar{\Lambda}\to p\pi^-\bar{p}\pi^+$ geometry diagram}
\end{figure}

The distribution of $\alpha$ depends on both the azimuth angles, $\phi_{\pi^\pm}$, of the momentum of $\pi^-$ and $\pi^+$ in the helicity rest frame of their own parent particles. As can be seen in the Fig. \ref{fig:plane}, the coordinate system is defined as following \cite{spin}. We choose the direction of the momentum of $\Lambda$ as the direction of $\hat{k}$ in the $\Lambda$'s helicity rest frame, that is to say, $\hat{k}=\hat{p}$. As for $\hat{\jmath}$ and $\hat{\imath}$, we have $\hat{\jmath}=\hat{k_0}\times \hat{k},\hat{\imath}=\hat{\jmath}\times \hat{k}$, where $\hat{k_0}$ is the direction of z-axis in the $J/\psi$ rest frame. In the $J/\psi$ rest frame, $\Lambda\bar{\Lambda}$ emit in the opposite direction, so  $\hat{p}_{\Lambda}=-\hat{p}_{\bar{\Lambda}}$. Then $\hat{\imath}(\Lambda)=\hat{\imath}(\bar{\Lambda})$, $\hat{\jmath}(\Lambda)=-\hat{\jmath}(\bar{\Lambda})$ and $\hat{k}(\Lambda)=-\hat{k}(\bar{\Lambda})$. 

In the $J/\psi$ rest frame, $\alpha$ equals the difference of $\phi_{\pi^+}$ and $\phi_{\pi^-}$. But in the $\Lambda\bar{\Lambda}$ helicity rest frame ,$\hat{\imath}(\Lambda)$, $\hat{\imath}(\bar{\Lambda})$ are in the same direction and $\hat{k}(\Lambda)$,$\hat{k}(\bar{\Lambda})$ are in the opposite direction. On this condition, $\alpha$ equals the sum of $\phi_{\pi^+}$ and $\phi_{\pi^-}$. Due to $\alpha\in[0,\pi]$, the definition can be written as following.

\begin{equation}\label{equ:alpha}
\alpha=
\begin{cases}
\phi_{\pi^-}+\phi_{\pi^+} & \phi_{\pi^-}+\phi_{\pi^+}\in[0,\pi]\\
2\pi-\phi_{\pi^-}-\phi_{\pi^+} & \phi_{\pi^-}+\phi_{\pi^+}\in(\pi,2\pi] \\
\phi_{\pi^-}+\phi_{\pi^+}-2\pi & \phi_{\pi^-}+\phi_{\pi^+}\in(2\pi,3\pi]\\
4\pi-\phi_{\pi^-}-\phi_{\pi^+}& \phi_{\pi^-}+\phi_{\pi^+}\in(3\pi,4\pi] 
 \\
\end{cases}
\end{equation}
The distribution of $\alpha$ can be expressed by  
\begin{equation}\label{equ:theory}
\begin{aligned}
W(\alpha)=\int_{0}^{2\pi}W^\prime(\phi_{\pi^-},\phi_{\pi^+})d\phi_{\pi-}\\=\int_{0}^{\alpha}W'(\phi_{\pi^-},\alpha-\phi_{\pi^-})d\phi_{\pi^-}\\+\int_{0}^{2\pi-\alpha}W'(\phi_{\pi^-},2\pi-\phi_{\pi^-}-\alpha)d\phi_{\pi^-}\\+\int_\alpha^{2\pi}W'(\phi_{\pi^-},\alpha+2\pi-\phi_{\pi^-})d\phi_{\pi^-}\\+\int_{2\pi-\alpha}^{2\pi} W'(\phi_{\pi^-},4\pi-\alpha-\phi_{\pi^-})d\phi_{\pi^-}
\end{aligned}
\end{equation}

where $W'(\phi_1,\phi_2)$ is the probability density function of $\phi_{\pi^-}=\phi_1$ and $\phi_{\pi^+}=\phi_2$ simultaneously.


Then, the angular distribution of the decay chain $ J/\psi\to\Lambda\bar{\Lambda}\to p\pi^-\bar{p}\pi^+ $ is \cite{spin}
\begin{equation}\label{equ:Joint}
\begin{aligned}
W''(\theta_{\pi^-},\phi_{\pi^-};\theta_{\pi^+},\phi_{\pi^+})=\\\frac{1}{4\pi}\sum_{l_1\geqslant0}^{2J_1}\sum_{l_2\geqslant0}^{2J_2}C_{l_1}(0,0;0,0)C_{l_2}(0,0;0,0)
\times\\ \sum_{m_1,m_2}t_{m_1,m_2}^{l_1,l_2}(C_1,C_2)Y_{l_1m_1}(\theta_1,\phi_1)Y_{l_2m_2}(\theta_2,\phi_2)
\end{aligned}
\end{equation}
where \cite{LandY}
\[C_0(0,0;0,0)=\frac{2}{\pi},C_1(0,0;0,0)=\frac{2a}{\pi}\]
W'' is the probability density function of $\theta_{\pi^-}$,$\phi_{\pi^-}$,$\theta_{\pi^+}$,$\phi_{\pi^+}$ and $t_{m_1,m_2}^{l_1,l_2}$ are the joint multipole parameters of $\Lambda\bar{\Lambda}$ system before the decay, which can be derived directly from the density matrix of initial state. $Y_{lm}$ are the spherical harmonic functions. 

The density matrices of initial state of $\Lambda$ and $\bar{\Lambda}$ rely on the initial polarization of $J/\psi$. If the initial polarization of $J/\psi$ was isotropic, the partition of the $J/\psi$ particles with the spin projection 1,0,-1 on z-axis in the helicity rest frame are all one third. The spin-parity of $J/\psi$ is $J^P=1^-$, parity conservation and angular momentum conservation requires that only S-wave and D-wave exist in the decay $J/\psi\to\Lambda\bar{\Lambda}$. However, the system of $\Lambda\bar{\Lambda}$ related by s-wave is in a pure state, which does not make contributions to entanglement. Under the circumstance that the initial spin projection on z-axis of $J/\psi$ is 1 or -1, the wave function before and after decay of d-wave can be written as 
\begin{equation}\label{equ:Jpsid}
\begin{aligned}
\chi_{J/\psi}|1,1>
=\sqrt{\frac{3}{5}}\mathbf{L}|2,2>\chi_{\Lambda\bar{\Lambda}}|1,-1>\\-\sqrt{\frac{3}{10}}\mathbf{L}|2,1>\chi_{\Lambda\bar{\Lambda}}|1,0>\\+\sqrt{\frac{1}{10}}\mathbf{L}|2,0>\chi_{\Lambda\bar{\Lambda}}|1,1>
\end{aligned}
\end{equation}
the joint density matrices of $\Lambda$ and $\bar{\Lambda}$ is

\begin{equation}\label{equ:initial}
\begin{aligned}
\rho^{\Lambda\bar{\Lambda}}=\frac{3}{5}|1,-1><1,-1|+\frac{3}{10}|1,0><1,0|+\frac{1}{10}|1,1><1,1|\\=\frac{3}{5}\left(\begin{array}{cc}0 & 0 \\0 & 1\end{array}\right)\otimes\left(\begin{array}{cc}1 & 0 \\0 & 0\end{array}\right)+\frac{1}{10}\left(\begin{array}{cc}1 & 0 \\0 & 0\end{array}\right)\otimes\left(\begin{array}{cc}0 & 0 \\0 & 1\end{array}\right)\\+\frac{3}{20}[\left(\begin{array}{cc}1 & 0 \\0 & 0\end{array}\right)\otimes\left(\begin{array}{cc}1 & 0 \\0 & 0\end{array}\right)+\left(\begin{array}{cc}0 & 0 \\0 & 1\end{array}\right)\otimes\left(\begin{array}{cc}0 & 0 \\0 & 1\end{array}\right)\\+\left(\begin{array}{cc}0 & 0 \\1 & 0\end{array}\right)\otimes\left(\begin{array}{cc}0 & 0 \\1 & 0\end{array}\right)+\left(\begin{array}{cc}0 & 1 \\0 & 0\end{array}\right)\otimes\left(\begin{array}{cc}0 & 1 \\0 & 0\end{array}\right)]
\end{aligned}
\end{equation}

where the terms before $\otimes$ are the density matrix of $\Lambda$ in $\Lambda$'s helicity rest frame. Likewise, the terms  after $\otimes$ are the density matrix of $\bar{\Lambda}$ in $\bar{\Lambda}$'s helicity rest frame. Afterwards, the joint multipole parameters of the initial states are 
\begin{equation}\label{equ:jointmultipole}
\begin{array}{|c|c|c|c|c|}
\hline t & t^{0\_}_{0\_} & t^{1\_}_{0\_} & t^{1\_}_{1\_} & t^{1\_}_{-1\_} \\
\hline t^{\_0}_{\_0} & 1 & -\frac{\sqrt{3}}{6} & -\frac{\sqrt{6}}{20} & \frac{\sqrt{6}}{20} \\
\hline t^{\_1}_{\_0} & \frac{\sqrt{3}}{6} & -\frac{2}{15} & 0 & 0 \\
\hline t^{\_1}_{\_1} & -\frac{\sqrt{6}}{20} & 0 & \frac{1}{10} &  0 \\
\hline t^{\_1}_{\_-1} & \frac{\sqrt{6}}{20} & 0 & 0 & \frac{1}{10} \\
\hline
\end{array}
\end{equation}
Then the distribution of $W(\phi_{\pi^-},\phi_{\pi^+})$ can be derived from Eq. (\ref{equ:Joint}) and Eq. (\ref{equ:jointmultipole})
\begin{equation}\label{equ:phi}
\begin{aligned}
W'(\phi_{\pi^-},\phi_{\pi^+})=\frac{1}{4\pi^2}+\frac{3}{20\pi^3}a(\cos\phi_{\pi^-}\\+\cos\phi_{\pi^+})+\frac{3}{10\pi^4}a^2\cos(\phi_{\pi^-}+\phi_{\pi^+})
\end{aligned}
\end{equation}

After combining equation Eq. (\ref{equ:theory}) and Eq. (\ref{equ:phi})together, and having $\theta_1 and \theta_2$ of integration, the result can be obtained, which is written as
\begin{equation}
W(\alpha)_{(s_z)_{J/\psi}=\pm 1}=\frac{1}{\pi}+\frac{6a^2\cos\alpha}{5\pi^3}
\end{equation} after normalization. $a=0.642$  \cite{PDG}, which is a constant determined by the reaction $\Lambda\to p\pi^-$.

For the other 1/3 of the decays in which $J/\psi$ particle has the spin projection on z-axis 0, the same procedure as above can be applied. The result distribution of $\alpha$ is
\begin{equation}
W(\alpha)_{(s_z)_{J/\psi}=0}=\frac{1}{\pi}+\frac{8a^2\cos\alpha}{5\pi^3}
\end{equation}

Then the final result of the distribution of $\alpha$ is
$\frac{1}{3}W(\alpha)_{(s_z)_{J/\psi}=0}+\frac{2}{3}W(\alpha)_{(s_z)_{J/\psi}=\pm 1}$

\begin{equation}\label{equ:finalresult}
W(\alpha)=\frac{1}{\pi}+\frac{4a^2\cos\alpha}{3\pi^3}
\end{equation}
     
If $\Lambda$ and $\bar{\Lambda}$ are not entangled, the decays of $\Lambda$ and $\bar{\Lambda}$ should be regarded as independent. So the Eq. (\ref{equ:theory}) turns into 
\begin{equation}\label{equ:independent}
\begin{aligned}
W(\alpha)=\int_{0}^{\alpha}W'(\phi_{\pi^-})W'(\alpha-\phi_{\pi^-})d\phi_{\pi^-}\\+\int_{0}^{2\pi-\alpha}W'(\phi_{\pi^-})W'(2\pi-\phi_{\pi^-}-\alpha)d\phi_{\pi^-}\\+\int_\alpha^{2\pi}W'(\phi_{\pi^-})W'(\alpha+2\pi-\phi_{\pi^-})d\phi_{\pi^-}\\+\int_{2\pi-\alpha}^{2\pi} W'(\phi_{\pi^-})W'(4\pi-\alpha-\phi_{\pi^-})d\phi_{\pi^-}
\end{aligned}
\end{equation}

The only bond between $\Lambda$ and $\bar{\Lambda}$ is the conservation of angular momentum and parity. 
According to the Ref. \cite{spin}, the angular distribution of the decay $\Lambda\to p\pi^-$ as well as $\bar{\Lambda}\to \bar{p}\pi^+$ is 

\begin{equation}\label{equ:inde_dis}
W'(\theta_\pi,\phi_\pi)=\frac{1}{\sqrt{4\pi}}\sum_{l=0}^{2J}C_{l}(0,0;0,0)\times\sum_m t^{l_*}_m Y_{lm}(\theta_\pi,\phi_\pi)
\end{equation}
Considering the initial state of $\Lambda$, the multipole parameters are 
\begin{equation}\label{equ:tparameter}
t^0_0=1,
t^0_1=\frac{1}{\sqrt{3}}\mathcal{P}_z,
t^1_1=-\frac{1}{\sqrt{6}}(\mathcal{P}_x+i\mathcal{P}_y),
t^{-1}_1=\frac{1}{\sqrt{6}}(\mathcal{P}_x-i\mathcal{P}_y)
\end{equation}

Then the distribution of $\theta_\pi,\phi_\pi$ can be derived from Eq. (\ref{equ:inde_dis}) 
and 
Eq. (\ref{equ:tparameter})
\begin{equation}\label{equ:final}
W'(\theta_\pi,\phi_\pi)=\frac{1}{2\pi}(1+a\frac{\mathcal{P}\cdot\mathbf{p_\pi}}{p_\pi})
\end{equation}

The distribution is only related to the initial state polarization vector below.
\begin{equation}\label{equ:polarization}
\mathcal{P}=(\sin\theta\cos\phi,\sin\theta\sin\phi,\cos\theta)^T
\end{equation}

Since $J/\psi$ is a particle with $J^P=1^-$ and both $\Lambda$ and $\bar{\Lambda}$ has spin $\frac{1}{2}$, the conservation of angular momentum requires that the spin of $\Lambda$ and $\bar{\Lambda}$ are in the same direction, which is 
\begin{equation}\label{equ:polarizaton}
\mathcal{P}_\Lambda=-\mathcal{P}_{\bar{\Lambda}}
\end{equation}
if written in their own helicity rest frame. 
The polarization of $\Lambda$ depends on the polarization of $J/\psi$, which is assumed to be isotropic. Thus, the polarization of $\Lambda$ and $\bar{\Lambda}$ can also be taken as isotropic. Combining the Eq. (\ref{equ:independent}) and Eq. (\ref{equ:final}) together, the final result of the distribution of $\alpha$ can be obtained, though the polarization vector of $\Lambda$ remains in it. As we discussed earlier, the polarization of $\Lambda$ is isotropic, which suggests uniform distribution after integrating the $\mathcal{P}_{\bar{\Lambda}}$ over $4\pi$ solid angle.
\begin{equation}\label{blabla}
W(\alpha)=\frac{1}{\pi}
\end{equation}

Under the condition of entanglement, the joint multipole parameters are involved to describe the initial state as the language of QM, which suggests that the entanglement of the two particles $\Lambda\bar{\Lambda}$ is considered since the initial state is mix instead of pure. Density matrix is used as the representation of QM instead of wave function only. Due to the fact that the parameter we choose to describe the entanglement, $\alpha$, is related to both sides of the decay while wave function is insufficient for a multi-particle system whose state can not be written as a wave function but a density matrix. So the two particles are viewed as a whole to compute the distribution under the case that QM is right. 

While without the consideration of entanglement, the angular distribution of the two decays are taken as independent. Only conservation of angular momentum is considered, and the initial polarization vector of the two particles are definite, though not known, which is the same as classic mechanics, so that the result can be taken as the prediction of HVTs.

This method can be universally used in all of the two-body decay chains $A\to C_1+C_2,C_1\to E_1+F_1,C_2\to E_2+F_2$ as long as the angle between two decay planes can be reconstructed in experiment, which is feasible to be carried out at experiments. Here we use $J/\psi\to\Lambda\bar{\Lambda}\to p \pi^-\to\bar{p}\pi^+$ to analysis. This decay channel can also be used to measure the CP violation\cite{cp}.

\begin{figure}[h]
\centering  
\includegraphics[width=9.5cm]{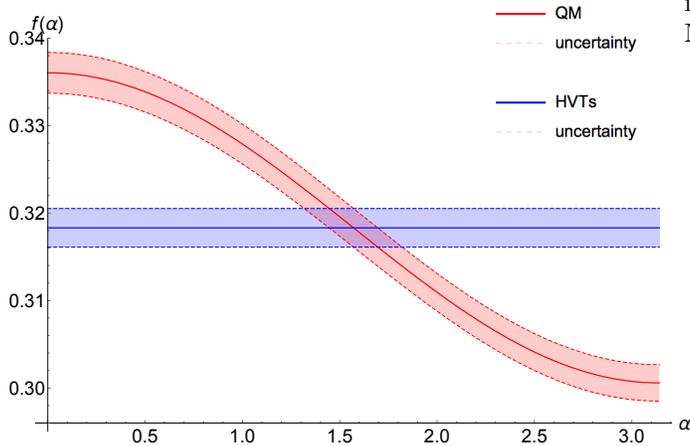}
\caption{\label{fig:deviation}%
uncertainty band of two distribution with 1.3 billion $J/\psi$ divided into 40 intervals}
\end{figure}

From the 
BES\uppercase\expandafter{\romannumeral3} detector at the BEPCII collider, about 1.3 billion $J/\psi$ events are collected and the branching ratio of $J/\psi\to\Lambda\bar{\Lambda}$ is about $1.6\times 10^{-3}$\cite{PDG}. Combining the efficiency of the detector which can be determined from MC simulation(here we take 40$\%$ as an approximation) , we can calculate the error band with 40 intervals, which is shown in the Fig. \ref{fig:deviation}.

Thus, the obvious discrepancy between the two distribution of $\alpha$ allows the experiments to test the two theories. Nonetheless, the concept of ``collapse'' is not mentioned here, which means that our method cannot test the ``locality'' but the ``reality'' questioned by the EPR paradox. However, the result from the experiment will indicate whether the $\Lambda$ and $\bar{\Lambda}$'s state violate reality or not and support one of QM and HVTs.

After all the discussion above, we may safely draw the conclusion that we have developed a new method to derive the distribution of an observable for discriminating QM and HVTs through experiment. The parameter we use, which is the angle between two decay planes, is accessible from the experiment and varies clearly as the results predicted by QM or HVTs. Therefore, the new method can be tested easily by experiment.

The authors thank Dr. Xun Chen for the valuable discussions. Our thanks also go to Bingtian Ye, who inspired the authors a lot with the method from quantum information. This work is supported by the National Natural Science Foundation of China(11661141008) 

\bibliographystyle{h-physrev}
\bibliography{entanglement}  
\end{document}